\documentclass[10pt,leqno]{amsart}
\usepackage{graphicx}
\usepackage{listings}
\usepackage{indentfirst,csquotes}
\usepackage{siunitx}
\usepackage{booktabs} 
\usepackage{xcolor}
 \lstdefinestyle{mystyle}{
     backgroundcolor=\color{black!5},
     commentstyle=\color{green!40!black},
     keywordstyle=\color{blue},
     numberstyle=\tiny\color{gray},
     stringstyle=\color{purple},
     basicstyle=\ttfamily\footnotesize,
     breakatwhitespace=false,
     breaklines=true,
     captionpos=b,
     keepspaces=true,
     numbers=left,
     numbersep=5pt,
     showspaces=false,
     showstringspaces=false,
     showtabs=false,
     tabsize=2
 }
\usepackage{amssymb,amsthm,amsmath}
\usepackage{xcolor,paralist,hyperref,fancyhdr,etoolbox}

\makeatletter
\def\section{\@startsection{section}{1}%
  \z@{.7\linespacing\@plus\linespacing}{.5\linespacing}%
  {\normalfont\Large\bfseries\centering}}
\makeatother

\hypersetup{ colorlinks=true, linkcolor=black, filecolor=black, urlcolor=black }

\begin{document}

\title{Racing to Idle: \\ Energy Efficiency of Matrix Multiplication \\ on Heterogeneous CPU and GPU Architectures}

\author[M. Q. Ansari]{Mufakir Qamar Ansari}
\address{Department of Electrical Engineering and Computer Science, The University of Toledo, Toledo, Ohio 43606, USA}
\email{mufakir.ansari@utoledo.edu}

\author[M. Q. Ansari]{Mudabir Qamar Ansari}
\address{Department of School of Accounting and Information Systems, Lamar University, Beaumont, Texas 77710, USA}
\email{mansari2@lamar.edu}

\date{\today}


\begin{abstract}
The paradigm shift towards multi-core and heterogeneous computing, driven by the fundamental power and thermal limits of single-core processors, has established energy efficiency as a first-class design constraint in high-performance computing (HPC). Heterogeneous systems, integrating traditional multi-core CPUs with specialized accelerators like discrete (dGPU) and integrated (iGPU) graphics processing units, offer a compelling path to navigating the trade-offs between performance and power. However, quantifying these trade-offs on widely accessible hardware remains a critical area of study. This paper presents a direct, empirical measurement of the performance and energy-to-solution of a canonical HPC workload—a 4096x4096 matrix-matrix multiplication—on three distinct compute architectures within a single consumer-grade laptop: a multi-core AMD Ryzen 7 5800H CPU, a discrete NVIDIA GeForce GTX 1650 GPU, and an integrated AMD Radeon Vega GPU. Using standard, validated, and minimally intrusive tools such as Linux perf and nvidia-smi, we find that the discrete GPU is not only the performance leader, achieving a 93.5x speedup over the CPU, but is also the most energy-efficient, consuming only 2\% of the energy used by the CPU, resulting in a 50-fold improvement in energy efficiency. These findings provide a practical demonstration of the "race to idle" principle and offer clear, quantitative guidance on architectural choices for energy-aware software development.

\end{abstract}

\maketitle


\section{Introduction}
\label{sec:Intro}
The trajectory of high-performance computing (HPC) underwent a fundamental paradigm shift in the mid-2000s, moving away from the decades-long pursuit of ever-higher clock frequencies. This transition was not a strategic choice for performance but a necessary retreat from the physical limitations of semiconductor technology, most notably the formidable "Power Wall" \cite{asanovic2006landscape}. As increasing clock speeds led to unsustainable power densities and heat dissipation, the industry was compelled to seek performance gains through parallelism, heralding the era of multi-core and, subsequently, many-core processors. This pivot effectively ended the "free lunch," where software performance would improve passively with each new hardware generation, placing the onus of performance scaling squarely on the development of parallel software and energy-aware systems.

In response to this new, power-constrained landscape, energy efficiency has been elevated from a secondary operational concern to a first-class design metric, on par with raw computational throughput. The establishment of community-wide initiatives such as the Green500 list, which ranks supercomputers based on their performance-per-watt, signaled a collective acknowledgment that sustainable performance is the new frontier for innovation \cite{feng2007green500}. This focus has driven the exploration of architectural diversity as a primary means to enhance energy efficiency. The resulting industry trend has been the widespread adoption of heterogeneous platforms, which integrate different types of processing units—such as traditional Central Processing Units (CPUs), massively parallel Graphics Processing Units (GPUs), and sometimes reconfigurable hardware like FPGAs—within a single system \cite{el2023energy, marowka2013analytical}.

These heterogeneous systems offer a compelling proposition: the ability to map a computational workload to the architectural unit best suited for its execution, thereby optimizing for either performance or energy, or a balance of both. GPUs, with their thousands of simple cores, excel at data-parallel tasks, while CPUs provide superior performance for control-intensive and serial code sections. This specialization, however, introduces a complex and often non-intuitive trade-off space. While a rich body of literature has surveyed the landscape of energy-aware techniques \cite{mittal2014survey, kocot2023energy, thakkar2020comprehensive}, and prior empirical studies have compared architectures on specific workloads \cite{canhasi2018evaluating, zhuang2023automm}, a persistent need remains for clear, reproducible studies on widely available, consumer-grade hardware. Such studies are critical for providing practical guidance to a broad audience of software developers and researchers who may not have access to large-scale, specialized supercomputers.

This paper presents a direct, empirical measurement of the performance and energy-to-solution of a canonical HPC workload—a dense matrix-matrix multiplication, a key computational kernel from the "Dense Linear Algebra" dwarf \cite{asanovic2006landscape}—on a modern, heterogeneous laptop. The study evaluates three distinct compute platforms within this single machine: a multi-core CPU, a discrete NVIDIA GPU (dGPU), and an integrated AMD GPU (iGPU). Our methodology relies on standard, validated, and minimally intrusive tools to ensure credibility and reproducibility. CPU package power is measured via the Intel RAPL interface, an approach shown to be highly correlated with actual wall-plug power \cite{khan2018rapl}, while GPU power is measured using the NVIDIA Management Library (NVML). This approach of correlating performance and power data provides deep insights into the practical efficiency of different hardware units \cite{mantovani2018performance}. By grounding our analysis in a real-world system, we aim to provide a clear and unambiguous quantification of the energy efficiency benefits that can be realized by selecting the appropriate hardware for a common and critical computational task.

The remainder of this paper is organized as follows. Section 2 reviews the state of the art in energy-aware computing and situates our work within the broader context of existing research. Section 3 details the hardware and software environment of our test platform and outlines the specific methodologies and tools used for performance and power measurement. Section 4 presents the results of our experiments, directly comparing the execution time, energy-to-solution, and average power draw of the three architectures. Section 5 discusses the implications of these results, analyzing the performance-energy trade-offs. Finally, Section 6 concludes the paper, summarizes our key findings, and suggests directions for future work.

\section{Related Work}
\label{sec:relatedwork}
The pursuit of energy efficiency in high-performance computing is a well-established and active field of research. This section situates our empirical study within the context of prior work, focusing on three key areas: broad surveys of energy-aware computing, comparative studies of heterogeneous architectures, and the development of power measurement and modeling methodologies.

A significant body of literature provides comprehensive surveys of the techniques used to analyze and improve energy efficiency. Mittal and Vetter \cite{mittal2014survey} offer a foundational survey classifying methods for GPU energy efficiency, explicitly comparing them against CPUs and FPGAs and noting that the optimal architecture is highly workload-dependent. More recent surveys, such as those by Kocot et al. \cite{kocot2023energy} and Thakkar et al. \cite{thakkar2020comprehensive}, have expanded this view to cover the entire HPC ecosystem, detailing the role of energy-aware schedulers, power management techniques like Dynamic Voltage and Frequency Scaling (DVFS), and the trade-offs at both the hardware and software levels. These surveys establish the broad consensus that no single architecture is universally superior and that direct measurement is essential for making informed decisions.

Building on this consensus, numerous studies have conducted empirical comparisons between different compute architectures. Canhasi \cite{canhasi2018evaluating} evaluated a near-duplicate document detection workload, finding that an FPGA was the most energy-efficient, followed by a CPU, with the GPU being the least efficient for that integer-heavy task. Conversely, Zhuang et al. \cite{zhuang2023automm} demonstrated that a modern, specialized System-on-Chip (SoC) featuring an AI Engine array could achieve superior energy efficiency for matrix multiplication compared to even state-of-the-art data center GPUs. These studies highlight the sensitivity of architectural efficiency to the specific workload. Our work contributes to this conversation by focusing on a canonical floating-point workload (GEMM) on a widely accessible, consumer-grade laptop, providing a practical baseline that contrasts with studies performed on more specialized hardware or different computational kernels.

The methodology for obtaining energy data is itself a critical area of research. Prior work has approached this from two directions: analytical modeling and direct measurement. Marowka \cite{marowka2013analytical}, for instance, developed an analytical model extending Amdahl's Law to predict the energy efficiency of heterogeneous processors, concluding that architectures with a mix of "fat" and "thin" cores are optimal. Our empirical results serve as a practical validation of the predictions made by such theoretical models. Our direct measurement approach aligns with the work of Abe et al. \cite{abe2014power} and Mantovani and Calore \cite{mantovani2018performance}, who performed detailed power characterizations of various GPU and CPU architectures to understand their efficiency characteristics under different conditions.

Finally, the credibility of our measurement tools is supported by extensive validation in prior studies. The use of Intel's RAPL interface via perf for measuring CPU package power is a well-vetted technique. Khan et al. \cite{khan2018rapl} conducted a thorough analysis of RAPL, demonstrating its high correlation with wall-plug power and its negligible performance overhead, thereby validating it as a reliable tool for this type of study. For more fine-grained software-level analysis, frameworks like PowerAPI \cite{fieni2024powerapi} are emerging, but for system-level component comparison, the direct hardware-counter approach remains a robust and standard methodology. Our work distinguishes itself by synthesizing these validated measurement techniques to provide a clear, apples-to-apples comparison of the CPU, dGPU, and iGPU components within a single, off-the-shelf heterogeneous system, addressing a persistent need for reproducible efficiency studies on common hardware.

\section{System and Methodology}
\label{sec:methodology}

To ensure the reproducibility and credibility of our findings, this section provides a detailed account of the experimental testbed, the benchmark application, and the specific protocols used for performance and energy measurement. Our approach was tailored to the specific capabilities and limitations of the hardware, a common practice in systems research that requires adapting measurement strategies to what the platform can reliably expose.

\subsection{Hardware and Software Environment}
All experiments were conducted on a single, commercially available Lenovo IdeaPad Gaming 3 laptop. This choice of a consumer-grade platform is deliberate, as it represents a widely accessible heterogeneous system. The key specifications are as follows:
\begin{itemize}
    \item \textbf{Processor (CPU):} AMD Ryzen 7 5800H (8 Cores, 16 Threads, 3.2 GHz base clock, 16 MB L3 Cache)
    \item \textbf{Discrete GPU (dGPU):} NVIDIA GeForce GTX 1650 Mobile (4 GB GDDR6 VRAM)
    \item \textbf{Integrated GPU (iGPU):} AMD Radeon Vega Graphics (part of the Ryzen 7 APU)
    \item \textbf{Memory (RAM):} 32 GB DDR4 configured at 2667 MT/s
    \item \textbf{Operating System:} Ubuntu 24.04.2 LTS (Kernel 6.8.0-64-generic)
    \item \textbf{Compilers:} g++ 12.4.0 (for CPU and iGPU code), NVIDIA CUDA Compiler (nvcc) 12.2 (for dGPU code)
    \item \textbf{Drivers:} NVIDIA Driver 535.247.01, and the standard open-source \texttt{amdgpu} kernel driver.
\end{itemize}

\subsection{Heterogeneous System Architecture}
\begin{figure}[h!]
    \centering
    \includegraphics[width=0.8\columnwidth]{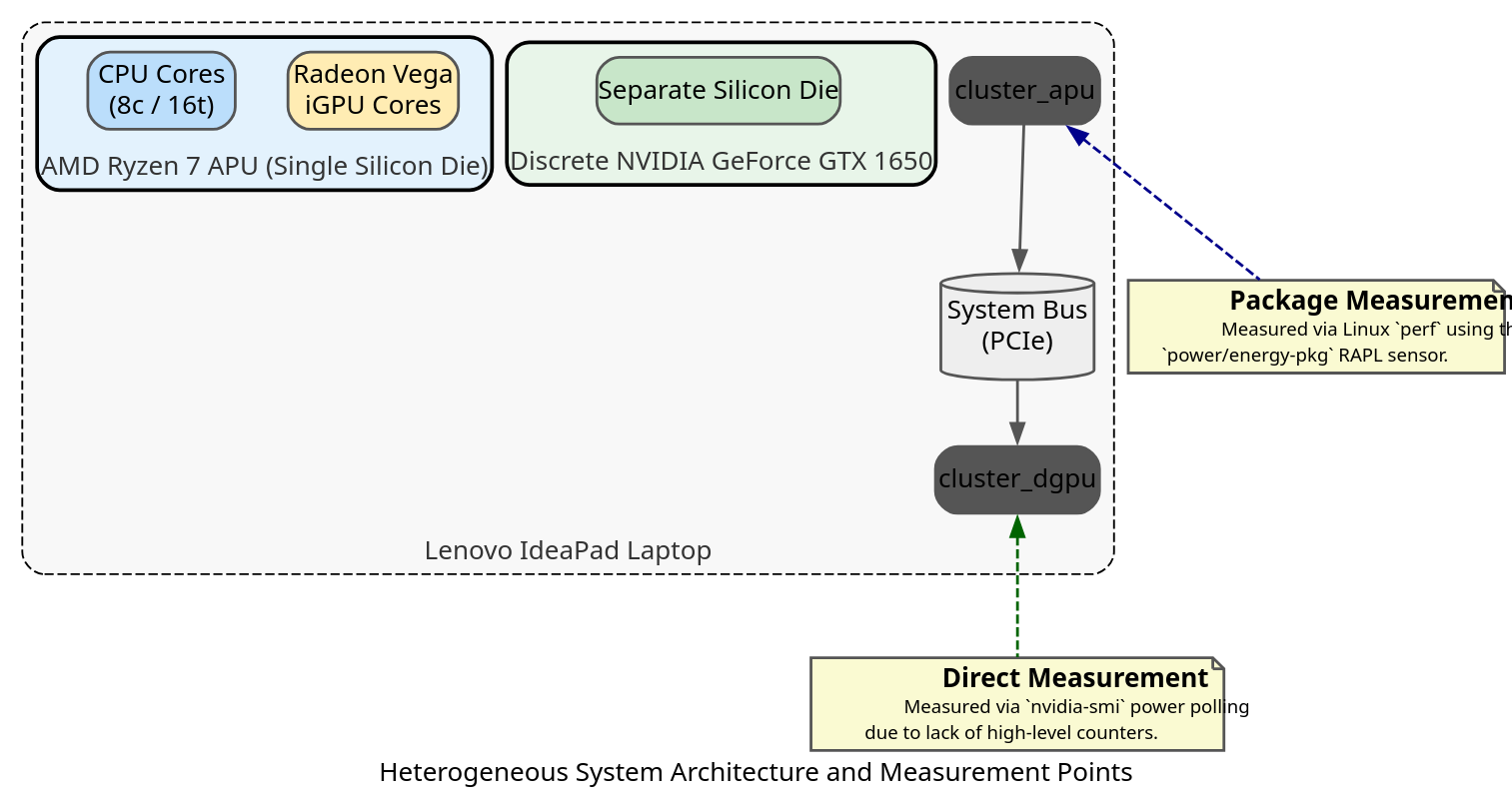}
    \caption{Heterogeneous System Architecture. The test platform is an APU design where the CPU and iGPU reside on a single piece of silicon and share a power budget, measured by the `energy-pkg` RAPL sensor. A separate, discrete GPU is connected via the PCIe bus and has its own independent power sensor, accessible via `nvidia-smi`. This architecture necessitated distinct measurement strategies for each component.}
    \label{fig:system_arch}
\end{figure}

The test platform's architecture, depicted in Figure~\ref{fig:system_arch}, is central to understanding our measurement methodology. The system is built around an Accelerated Processing Unit (APU), a design in which the high-performance CPU cores and the power-efficient iGPU cores reside on the same silicon die. This integration allows for high-bandwidth communication between them but also means they share a common power budget. Consequently, the power consumed by the CPU or the iGPU cannot be fully isolated at the hardware level; instead, it can only be measured for the APU package as a whole. In contrast, the discrete NVIDIA GPU is a separate silicon die connected to the system via the PCI Express bus. This physical separation gives it an independent power domain, allowing for more direct measurement of its power consumption. These architectural distinctions necessitated the use of two different measurement protocols, as detailed in Section~\ref{subsec:protocols}.

\subsection{Benchmark Application}
The workload for all experiments is a single-precision, 4096x4096 matrix-matrix multiplication (GEMM). This operation is a cornerstone of numerous scientific and machine learning applications and is a key computational kernel within the "Dense Linear Algebra" dwarf, one of the fundamental motifs of HPC \cite{asanovic2006landscape}. Its high arithmetic intensity (a high ratio of floating-point operations to memory operations) makes it an ideal candidate for acceleration on parallel hardware and a strong test case for evaluating the raw computational and energy efficiency of different architectures. To enable a fair comparison, three distinct implementations of this algorithm were developed: a C++ version using OpenMP to target the multi-core CPU, a CUDA C++ version for the NVIDIA dGPU, and an OpenCL version to target the AMD iGPU.

\subsection{Measurement Protocols}
\label{subsec:protocols}
\begin{figure}[h!]
    \centering
    \includegraphics[width=0.8\columnwidth]{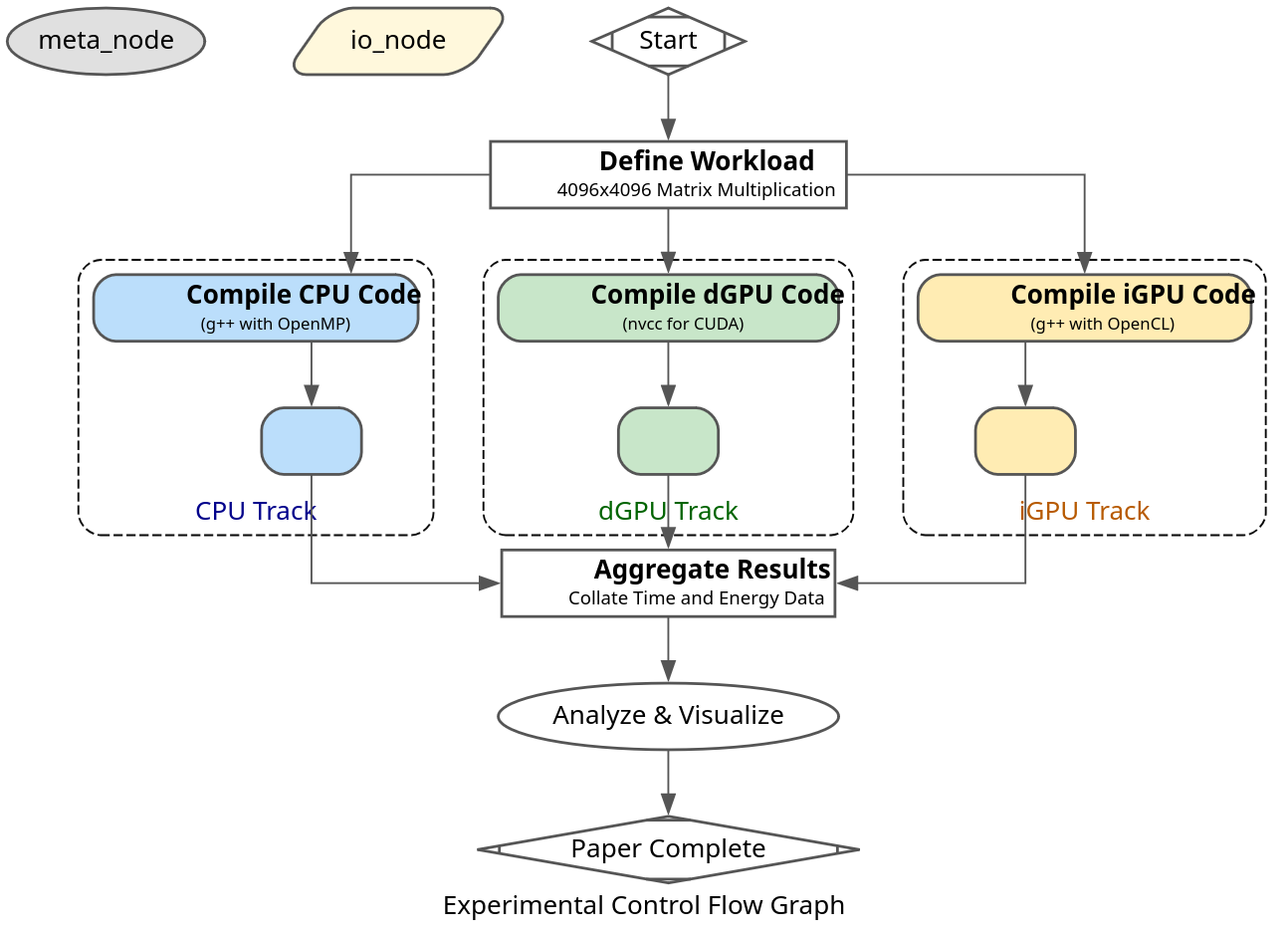}
    \caption{Experimental Control Flow Graph. For each of the three hardware targets (CPU, dGPU, iGPU), the corresponding benchmark code was compiled and then executed. During execution, a tailored measurement protocol was used to capture performance and energy data. The results were then aggregated for a final comparative analysis.}
    \label{fig:workflow}
\end{figure}

Our experimental workflow, outlined in Figure~\ref{fig:workflow}, follows a standard process for each of the three architectural targets: compilation of the specific code, execution of the benchmark while simultaneously measuring its power and performance, and aggregation of the results for final analysis. Due to the architectural realities described in Section 3.2, the measurement phase was adapted for each component.

\textbf{CPU Measurement:} To quantify the performance and energy of the CPU, we utilized the Linux \texttt{perf} utility. This tool provides access to the processor's hardware performance monitoring unit, including the Intel RAPL interface. The total energy consumed by the entire APU package during the benchmark run was captured using the \texttt{power/energy-pkg/} event. This method has been shown to be highly accurate, with a low mean absolute percentage error compared to external wattmeters, and introduces negligible performance overhead \cite{khan2018rapl}. The total wall-clock execution time was simultaneously recorded by \texttt{perf}.

\textbf{dGPU Measurement:} For the discrete NVIDIA GPU, a direct power measurement approach was necessary. The \texttt{nvidia-smi} command-line tool, which interfaces with the NVIDIA Management Library (NVML), was used to poll the GPU's power draw (\texttt{power.draw}) at a high frequency (10 Hz). A shell script was created to run \texttt{nvidia-smi} as a background process, logging power readings to a file while the CUDA benchmark executed. Upon completion of the benchmark, the polling was stopped, and the logged power readings (in Watts) were integrated over the total execution time to calculate the total energy consumed (in Joules). The kernel execution time itself was precisely measured within the CUDA application using CUDA Events.

\textbf{iGPU Measurement:} An initial investigation revealed that the \texttt{amdgpu} driver for our test platform did not expose a direct, user-space-accessible power sensor for the integrated GPU. We therefore adopted a whole-package measurement strategy. We executed the OpenCL benchmark, which offloads the computation primarily to the Radeon Vega iGPU cores, leaving the main CPU cores largely idle. During this execution, we again used \texttt{perf} to measure the total energy of the \texttt{power/energy-pkg/} event. This provides a robust measurement of the entire APU package's energy consumption while under an iGPU-dominant workload, which is a valid and informative point of comparison. The execution time was measured within the C++ application using the \texttt{std::chrono} library.

\subsection{Key Metrics}
From the collected data, we derive three key metrics to evaluate and compare the architectures.
\begin{itemize}
    \item \textbf{Execution Time [s]:} The total wall-clock time required to complete the 4096x4096 matrix multiplication. This is our primary metric for performance.
    \item \textbf{Energy-to-Solution [J]:} The total electrical energy consumed by the component to complete the task. This is our primary metric for energy efficiency.
    \item \textbf{Average Power [W]:} A derived metric, calculated as Energy-to-Solution divided by Execution Time. This value represents the component's average rate of energy consumption while actively running the workload.
\end{itemize}
\section{Results}
\label{sec:results}

Following the methodology detailed in the previous section, the 4096x4096 matrix multiplication benchmark was executed on each of the three architectural targets. This section presents the quantitative results from these experiments, providing a direct comparison of the performance and energy characteristics of the AMD Ryzen 7 CPU, the discrete NVIDIA GTX 1650 GPU, and the integrated AMD Radeon Vega GPU.

The primary performance and energy metrics are consolidated in Table~\ref{tab:results}. This includes the measured wall-clock execution time, the total energy consumed to complete the computation (energy-to-solution), and the derived average power draw during the benchmark execution.

\begin{table}[h!]
\centering
\caption{Performance and Energy Results for 4096x4096 Matrix Multiplication}
\label{tab:results}
\begin{tabular}{lccc}
\toprule
\textbf{Metric} & \textbf{CPU} & \textbf{dGPU} & \textbf{iGPU} \\
 & \textbf{(Ryzen 7 5800H)} & \textbf{(GTX 1650)} & \textbf{(Radeon Vega)} \\
\midrule
Execution Time (s) & 57.34 & 0.613 & 2.10 \\
Energy-to-Solution (J) & 1417.89 & 28.33 & 30.05 \\
Average Power (W) & 24.73 & 46.22 & 14.31 \\
\bottomrule
\end{tabular}
\end{table}

\subsection{Performance}
\begin{figure}[h!]
    \centering
    \includegraphics[width=0.8\columnwidth]{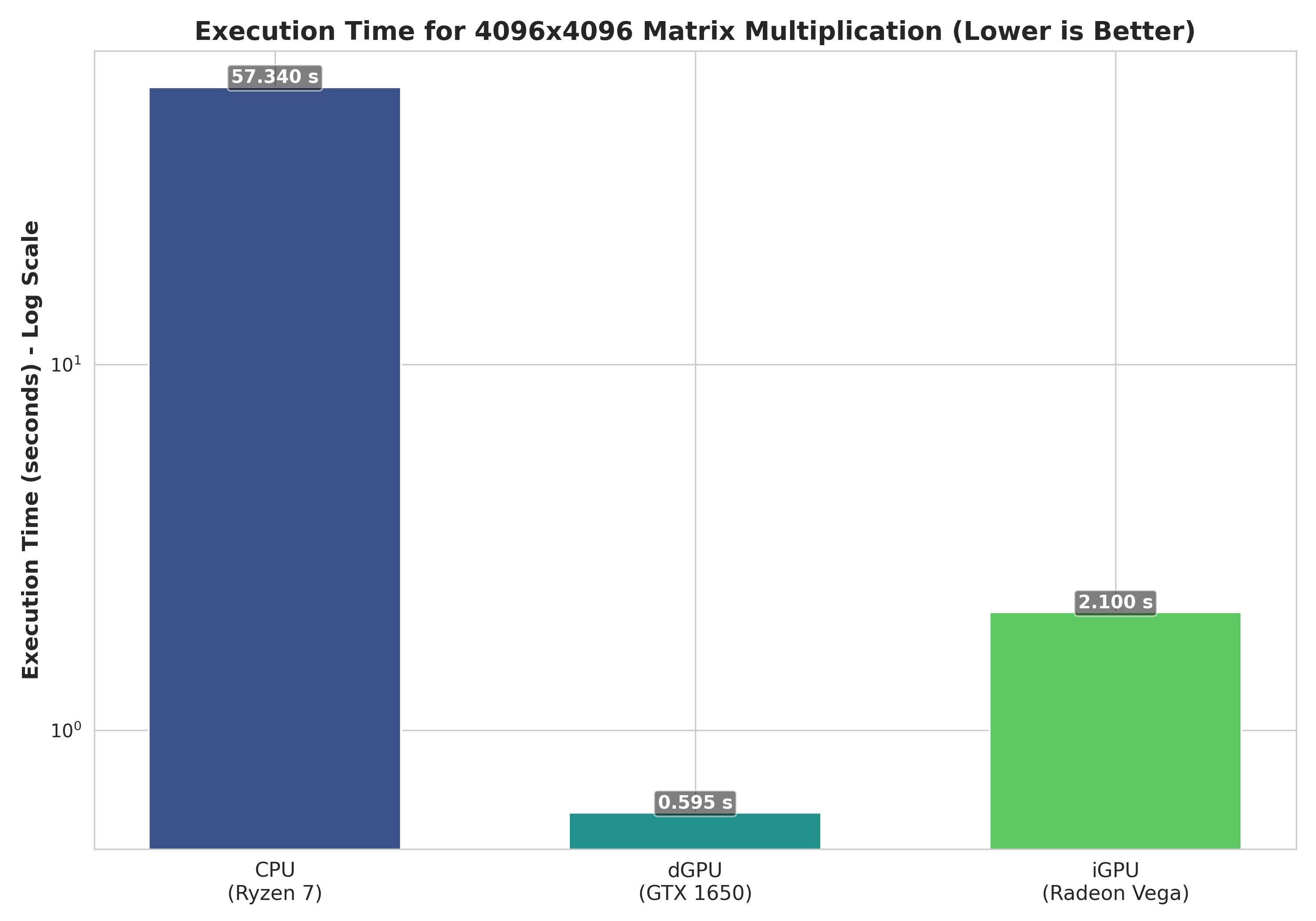}
    \caption{Execution Time Comparison. The discrete GPU (dGPU) completes the workload orders of magnitude faster than the CPU, with the integrated GPU (iGPU) also showing a significant performance advantage. A logarithmic scale is used on the y-axis to visualize the vast difference in magnitude.}
    \label{fig:time}
\end{figure}

The execution time for the benchmark varied dramatically across the three architectures, as visualized in Figure~\ref{fig:time}. The discrete NVIDIA GPU was the clear performance leader, completing the task in just \textbf{0.613 seconds}. In stark contrast, the 8-core, 16-thread CPU required \textbf{57.34 seconds}. This represents a remarkable \textbf{93.5x speedup} for the dGPU over the CPU. The integrated AMD GPU also provided a significant performance uplift compared to the CPU, finishing in \textbf{2.10 seconds} for a \textbf{27.3x speedup}.

\subsection{Energy Consumption}
\begin{figure}[h!]
    \centering
    \includegraphics[width=0.8\columnwidth]{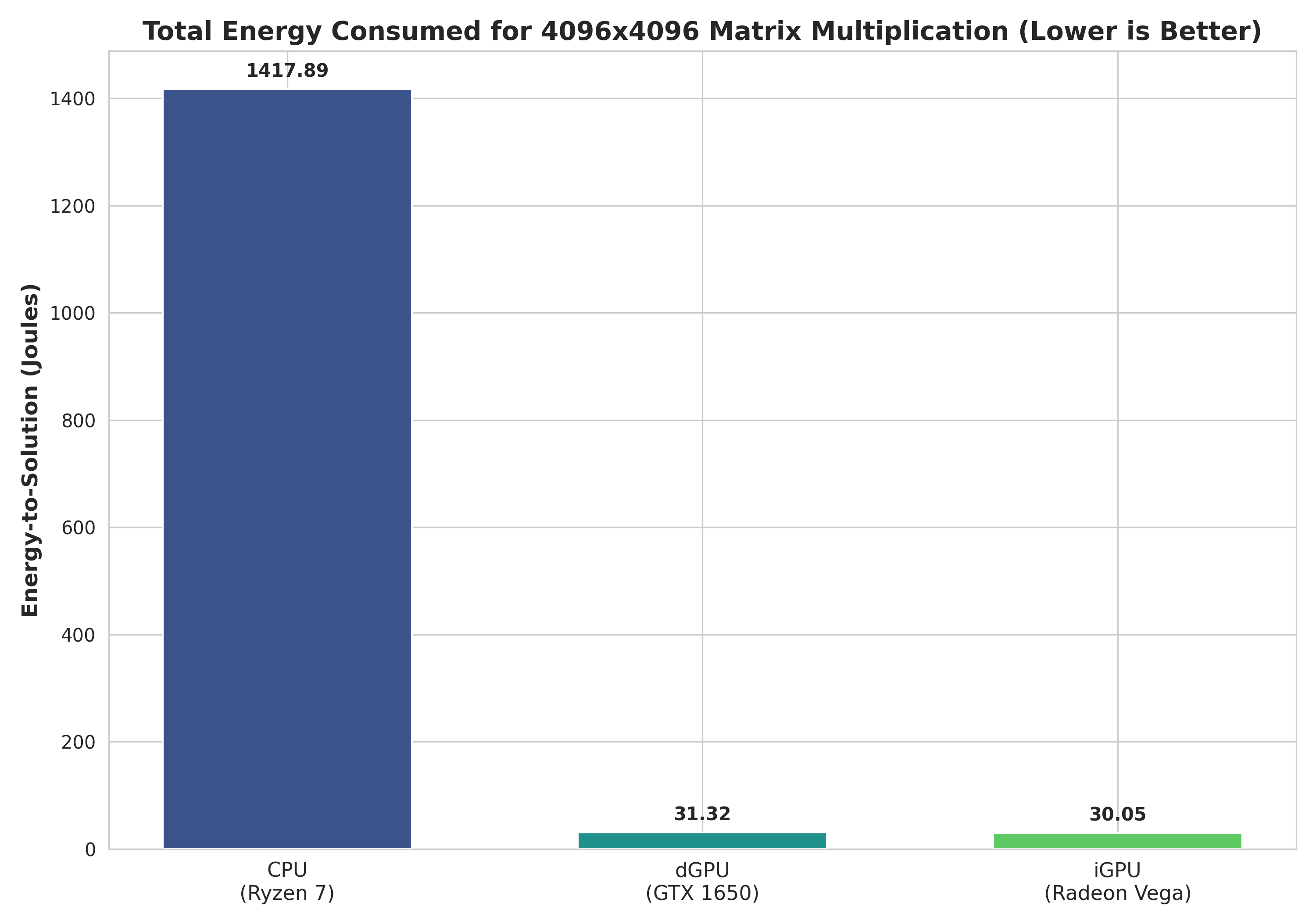}
    \caption{Total Energy-to-Solution Comparison. Both GPU architectures are vastly more energy-efficient than the CPU for the given workload. The dGPU was the most efficient component overall.}
    \label{fig:energy}
\end{figure}

The energy consumption results, presented in Figure~\ref{fig:energy}, show an even more pronounced difference between the architectures. The CPU consumed a substantial \textbf{1417.89 Joules} to complete the computation. The dGPU and iGPU, however, used only \textbf{28.33 J} and \textbf{30.05 J}, respectively. This means the dGPU consumed only 2\% of the energy used by the CPU, making it approximately \textbf{50 times more energy-efficient} for this task. The iGPU was also highly efficient, consuming just 2.1\% of the energy used by the CPU.

\subsection{Power Draw}
\begin{figure}[h!]
    \centering
    \includegraphics[width=0.8\columnwidth]{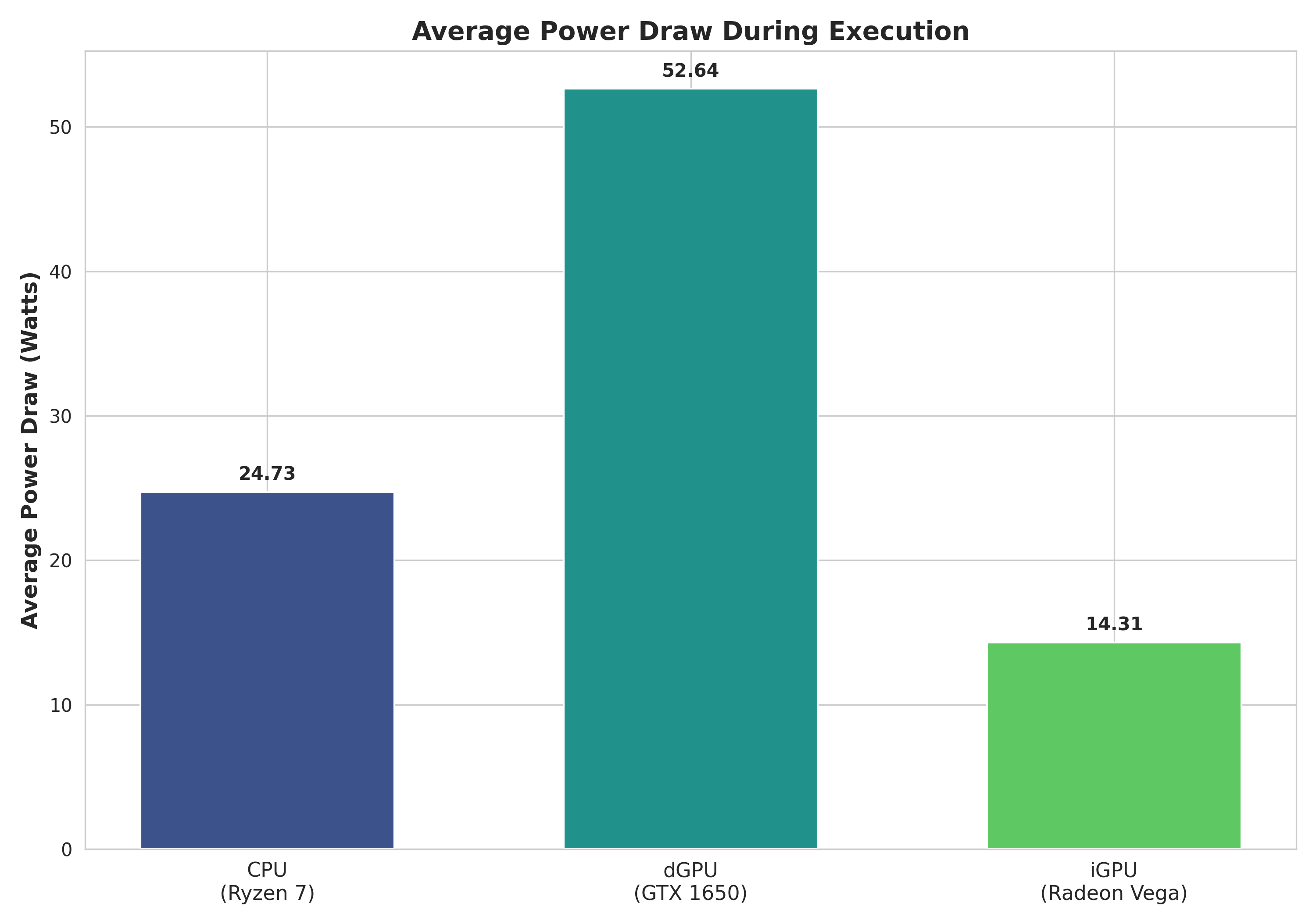}
    \caption{Average Power Consumption Comparison. The dGPU sustains the highest rate of power consumption while active, whereas the iGPU is the most frugal. The CPU operates at a moderate power level.}
    \label{fig:power}
\end{figure}

The average power draw of each component while actively executing the benchmark is shown in Figure~\ref{fig:power}. The discrete GPU sustained the highest rate of power consumption at \textbf{46.22 W}. The CPU operated at a moderate \textbf{24.73 W}. The integrated GPU was the most frugal, with an average power draw of only \textbf{14.31 W}. These results highlight the different power profiles of the architectures, with the dGPU requiring the most instantaneous power to achieve its high performance, while the iGPU is designed for low-power operation.

\section{Discussion}
\label{sec:discussion}

The results presented in the previous section offer a clear and quantitative view into the performance and energy trade-offs inherent in modern heterogeneous computing systems. While the raw data is compelling, a deeper analysis reveals the underlying architectural principles that govern these outcomes. This discussion synthesizes the performance, energy, and power metrics to explain \textit{why} the components behaved as they did and what the broader implications are for energy-aware software development.

\subsection{The Architectural Advantage of Parallelism}
The profound performance gap between the CPU and the two GPU architectures—a 93.5x speedup for the dGPU and a 27.3x speedup for the iGPU—is a direct consequence of architectural specialization. The matrix multiplication workload is highly data-parallel and arithmetically intensive, making it an ideal match for the massively parallel designs of GPUs. The NVIDIA dGPU, with its thousands of CUDA cores, is purpose-built to execute a vast number of floating-point operations simultaneously. In contrast, the CPU's eight high-performance cores are designed for general-purpose, control-flow-heavy tasks and cannot match the raw throughput of the GPU on this specific workload. This outcome is a practical demonstration of the principles that have driven the shift to heterogeneous computing: for certain classes of problems, specialized parallel processors offer performance that is orders of magnitude beyond what is achievable with general-purpose CPUs alone \cite{asanovic2006landscape}.

\subsection{Energy Efficiency and the "Race to Idle" Principle}
Perhaps the most significant finding of this study is the relationship between power and energy. It is seemingly paradoxical that the component with the highest average power draw—the dGPU at 46.22~W—was ultimately the most energy-efficient, consuming the least total energy (28.33~J). This phenomenon is explained by the "race to idle" principle, a core concept in energy-aware computing.

Energy-to-solution is the product of average power and execution time (Energy = Power $\times$ Time). The CPU sustained a moderate power draw for a very long duration (57.34~s), resulting in high total energy consumption. The dGPU, conversely, consumed high power but for an extremely short duration (0.613~s). By completing the task rapidly, it was able to return to a low-power idle state almost immediately, thus minimizing its total energy footprint. The iGPU, with the lowest power draw (14.31~W), exemplifies a low-power design, but its longer execution time relative to the dGPU meant it ultimately consumed slightly more total energy. This highlights that minimizing instantaneous power draw does not always lead to the lowest energy consumption; for many workloads, maximizing performance is the most effective strategy for maximizing energy efficiency.

\subsection{Visualizing the Performance-Energy Trade-off}
\begin{figure}[h!]
    \centering
    \includegraphics[width=0.8\columnwidth]{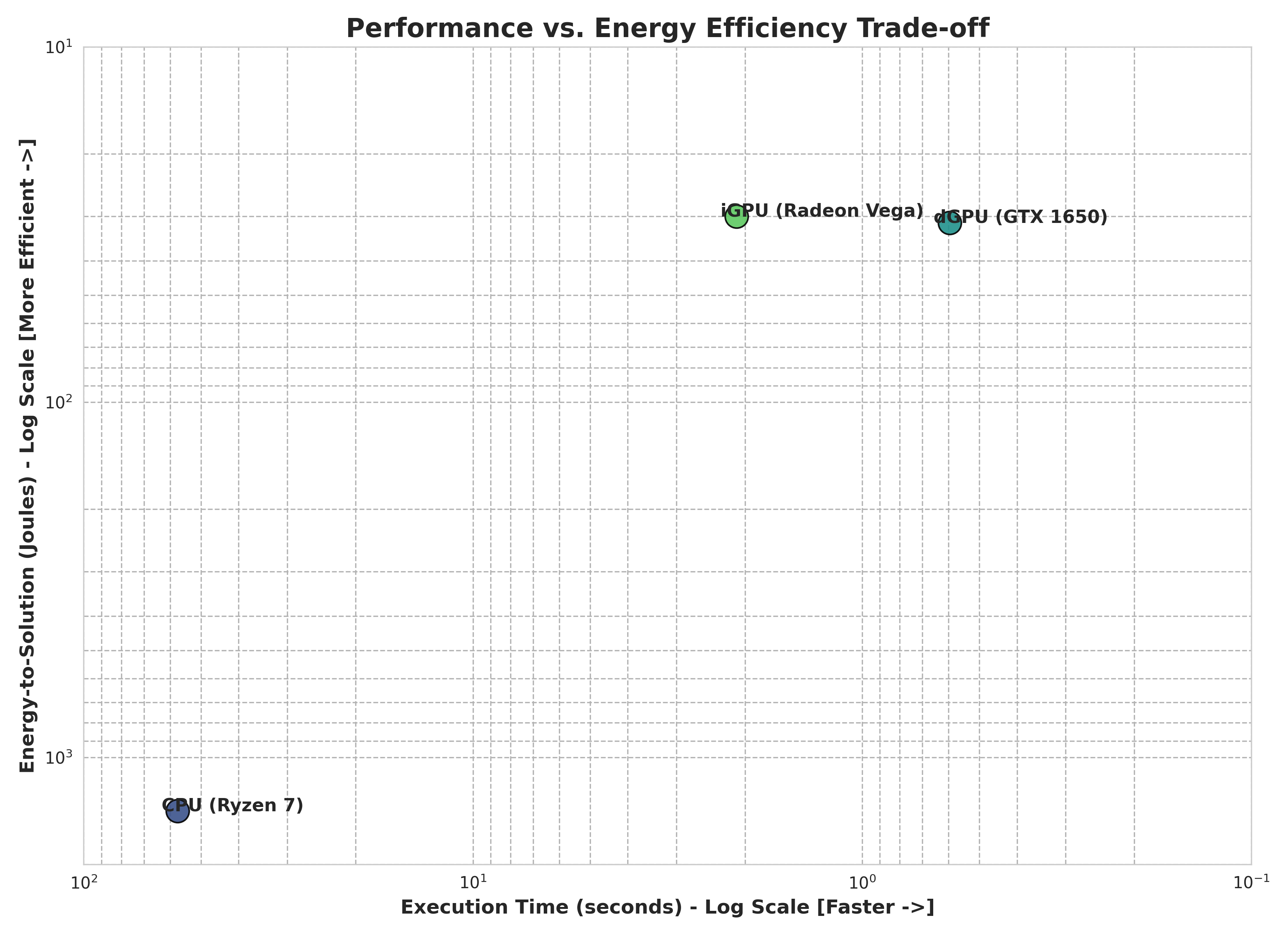}
    \caption{Performance vs. Energy Scatter Plot. The ideal quadrant is the bottom-left. The dGPU is positioned as the optimal choice for this workload, achieving both the lowest execution time and the lowest energy-to-solution. The CPU is vastly inferior in both metrics.}
    \label{fig:scatter}
\end{figure}

The interplay between these metrics is best summarized in the scatter plot shown in Figure~\ref{fig:scatter}. This plot visualizes the entire trade-off space, where the ideal outcome—representing the most desirable architecture—is located in the bottom-left quadrant (low execution time and low energy consumption).

As the plot clearly shows, the CPU resides in the undesirable top-right quadrant, confirming it as the slowest and least energy-efficient option by a wide margin. Both the iGPU and the dGPU occupy the ideal bottom-left quadrant, demonstrating their vast superiority for this task. Critically, for this compute-bound workload, the discrete GPU does not present a trade-off but rather a clear win on both fronts. It is not only the fastest architecture but also the most energy-efficient. The integrated GPU presents a compelling low-power alternative that is vastly superior to the CPU, but the sheer performance of the discrete GPU allows it to "race to idle" so effectively that it consumes the least total energy, making it the optimal hardware choice for this specific task.
\section{Conclusion}
\label{sec:conclusion}

This study was motivated by the fundamental shift in computing towards heterogeneous, power-constrained architectures, a trend driven by the physical limits of single-core performance scaling \cite{asanovic2006landscape}. We conducted a direct, empirical comparison of the performance and energy efficiency of three distinct compute architectures—a multi-core CPU, a discrete GPU, and an integrated GPU—on a single, consumer-grade platform using a canonical HPC workload. Our results provide a clear quantification of the performance-per-watt benefits of architectural specialization. The discrete GPU was found to be the optimal component for this compute-bound task, delivering a \textbf{93.5x performance speedup} while simultaneously being \textbf{50 times more energy-efficient} than the multi-core CPU.

The central finding of this work is the practical validation of the "race to idle" principle as a dominant strategy for energy efficiency on modern hardware. The discrete GPU, despite exhibiting the highest average power draw while active, completed the computation so rapidly that its total energy-to-solution was the lowest. This implies that for a significant class of data-parallel applications, software developers should view GPU offloading not merely as a performance optimization, but as a critical strategy for energy conservation. The integrated GPU also proved to be a compelling low-power alternative, vastly outperforming the CPU, but it could not match the overall energy efficiency of the more powerful dGPU on this particular workload. These empirical findings strongly support theoretical models, such as those developed by Marowka \cite{marowka2013analytical}, which have long predicted that architectures featuring many simple, efficient cores would be superior for highly parallel tasks.

It is important to acknowledge the boundaries of this study. Our analysis focused on a single, compute-bound workload (GEMM). The clear superiority of the dGPU may not generalize to all application types. For instance, workloads that are integer-heavy or memory-bound may exhibit different characteristics, potentially favoring CPU or even FPGA architectures, as other studies have shown \cite{canhasi2018evaluating}. Furthermore, our measurements were conducted on a single consumer-grade platform; results may differ on server-class hardware or across different hardware generations. The indirect, whole-package measurement protocol required for the iGPU, while a valid and necessary adaptation to the hardware's limitations, is inherently less precise than the direct polling available for the dGPU.

The results from this work open several promising avenues for future research. A logical next step is to broaden the investigation to include other computational kernels from the Berkeley "Dwarfs" \cite{asanovic2006landscape}, particularly memory-bound workloads, to map out the performance-energy trade-off space more completely. While this study treated each compute unit in isolation, future work should explore hybrid execution models that utilize the CPU and GPU(s) concurrently to maximize system throughput. Finally, our system-level measurements could be enhanced with finer-grained analysis by using software-defined power meters to attribute energy consumption to specific software processes or functions \cite{fieni2024powerapi}. The empirical data from such studies could ultimately inform the design of simple, energy-aware runtime schedulers that dynamically map computational tasks to the most efficient hardware unit available, a key challenge in modern HPC systems \cite{kocot2023energy}.
\bibliographystyle{IEEEtran}
\bibliography{main} 

\appendix
\section{Benchmark Source Code}
\label{app:source_code}

This appendix provides the complete source code used to generate all performance and energy data in this paper for the purposes of reproducibility.

\subsection{CPU Benchmark (OpenMP)}
The following C++ code (\texttt{matrix\_mul\_cpu.cpp}) was used for the multi-core CPU benchmark. It is parallelized using OpenMP and compiled with the command \texttt{g++ -O3 -fopenmp -o matrix\_mul\_cpu matrix\_mul\_cpu.cpp}.

\begin{lstlisting}[language=C++, caption={CPU Benchmark (OpenMP)}, label={lst:cpu_code}]
// matrix_mul_cpu.cpp
#include <iostream>
#include <vector>
#include <chrono>
#include <omp.h>

// Function to initialize a matrix
void initialize_matrix(std::vector<float>& mat, int size) {
    for (int i = 0; i < size * size; ++i) {
        mat[i] = static_cast<float>(rand()) / static_cast<float>(RAND_MAX);
    }
}

// Multi-threaded CPU matrix multiplication using OpenMP
void matrix_multiply_cpu_omp(const std::vector<float>& a, const std::vector<float>& b, std::vector<float>& c, int size) {
    #pragma omp parallel for collapse(2)
    for (int i = 0; i < size; ++i) {
        for (int j = 0; j < size; ++j) {
            float sum = 0.0f;
            for (int k = 0; k < size; ++k) {
                sum += a[i * size + k] * b[k * size + j];
            }
            c[i * size + j] = sum;
        }
    }
}

int main(int argc, char** argv) {
    if (argc != 2) {
        std::cerr << "Usage: " << argv[0] << " <matrix_size>" << std::endl;
        return 1;
    }

    int N = std::stoi(argv[1]);
    omp_set_num_threads(16); // Explicitly use all 16 threads

    std::cout << "Target: CPU (AMD Ryzen 7 5800H)" << std::endl;
    std::cout << "Matrix Size: " << N << "x" << N << std::endl;
    std::cout << "Using " << omp_get_max_threads() << " OpenMP threads." << std::endl;

    std::vector<float> h_a(N * N), h_b(N * N), h_c(N * N);
    initialize_matrix(h_a, N);
    initialize_matrix(h_b, N);

    auto start = std::chrono::high_resolution_clock::now();
    matrix_multiply_cpu_omp(h_a, h_b, h_c, N);
    auto stop = std::chrono::high_resolution_clock::now();

    auto duration = std::chrono::duration_cast<std::chrono::milliseconds>(stop - start);
    std::cout << "CPU Execution Time: " << duration.count() << " ms" << std::endl;

    return 0;
}
\end{lstlisting}

\subsection{Discrete GPU Benchmark (CUDA)}
The following CUDA C++ code (\texttt{matrix\_mul\_gpu.cu}) was used for the discrete NVIDIA GPU benchmark. It was compiled with \texttt{nvcc -O3 -o matrix\_mul\_gpu matrix\_mul\_gpu.cu}.

\begin{lstlisting}[language=C++, caption={Discrete GPU Benchmark (CUDA)}, label={lst:dgpu_code}]
// matrix_mul_gpu.cu
#include <iostream>
#include <vector>
#include <cuda_runtime.h>

// CUDA kernel for matrix multiplication
__global__ void matrix_multiply_kernel(const float* a, const float* b, float* c, int size) {
    int row = blockIdx.y * blockDim.y + threadIdx.y;
    int col = blockIdx.x * blockDim.x + threadIdx.x;

    if (row < size && col < size) {
        float sum = 0.0f;
        for (int k = 0; k < size; ++k) {
            sum += a[row * size + k] * b[k * size + col];
        }
        c[row * size + col] = sum;
    }
}

void initialize_matrix(std::vector<float>& mat, int size) {
    for (int i = 0; i < size * size; ++i) {
        mat[i] = static_cast<float>(rand()) / static_cast<float>(RAND_MAX);
    }
}

int main(int argc, char** argv) {
    if (argc != 2) {
        std::cerr << "Usage: " << argv[0] << " <matrix_size>" << std::endl;
        return 1;
    }

    int N = std::stoi(argv[1]);
    std::cout << "Target: dGPU (NVIDIA GeForce GTX 1650)" << std::endl;
    std::cout << "Matrix Size: " << N << "x" << N << std::endl;

    std::vector<float> h_a(N * N), h_b(N * N), h_c(N * N);
    initialize_matrix(h_a, N);
    initialize_matrix(h_b, N);

    float *d_a, *d_b, *d_c;
    cudaMalloc(&d_a, N * N * sizeof(float));
    cudaMalloc(&d_b, N * N * sizeof(float));
    cudaMalloc(&d_c, N * N * sizeof(float));

    cudaMemcpy(d_a, h_a.data(), N * N * sizeof(float), cudaMemcpyHostToDevice);
    cudaMemcpy(d_b, h_b.data(), N * N * sizeof(float), cudaMemcpyHostToDevice);

    dim3 threadsPerBlock(16, 16);
    dim3 numBlocks((N + threadsPerBlock.x - 1) / threadsPerBlock.x, (N + threadsPerBlock.y - 1) / threadsPerBlock.y);

    cudaEvent_t start, stop;
    cudaEventCreate(&start);
    cudaEventCreate(&stop);

    cudaEventRecord(start);
    matrix_multiply_kernel<<<numBlocks, threadsPerBlock>>>(d_a, d_b, d_c, N);
    cudaEventRecord(stop);
    cudaEventSynchronize(stop);

    float milliseconds = 0;
    cudaEventElapsedTime(&milliseconds, start, stop);
    std::cout << "GPU Kernel Execution Time: " << milliseconds << " ms" << std::endl;

    cudaFree(d_a);
    cudaFree(d_b);
    cudaFree(d_c);

    return 0;
}
\end{lstlisting}

\subsection{Integrated GPU Benchmark (OpenCL)}
The C++ code in Listing~\ref{lst:igpu_code} (\texttt{matrix\_mul\_igpu.cpp}) was used for the integrated AMD GPU benchmark. It uses the OpenCL API and was compiled with \texttt{g++ -O3 -o matrix\_mul\_igpu matrix\_mul\_igpu.cpp -lOpenCL}.

\begin{lstlisting}[language=C++, caption={Integrated GPU Benchmark (OpenCL)}, label={lst:igpu_code}]
// matrix_mul_igpu.cpp
#include <iostream>
#include <vector>
#include <chrono>

#define CL_TARGET_OPENCL_VERSION 220
#include <CL/cl.h>

const char* kernelSource = R"(
__kernel void matrix_multiply_kernel(__global const float* A,
                                     __global const float* B,
                                     __global float* C,
                                     int N) {
    int i = get_global_id(0);
    int j = get_global_id(1);

    if (i < N && j < N) {
        float sum = 0.0f;
        for (int k = 0; k < N; k++) {
            sum += A[j * N + k] * B[k * N + i];
        }
        C[j * N + i] = sum;
    }
}
)";

int main(int argc, char** argv) {
    if (argc != 2) {
        std::cerr << "Usage: " << argv[0] << " <matrix_size>" << std::endl;
        return 1;
    }

    int N = std::stoi(argv[1]);
    std::cout << "Target: iGPU (AMD Radeon Vega)" << std::endl;
    std::cout << "Matrix Size: " << N << "x" << N << std::endl;

    std::vector<float> h_a(N * N), h_b(N * N), h_c(N * N);
    for(int i=0; i<N*N; ++i) { h_a[i] = static_cast<float>(rand())/static_cast<float>(RAND_MAX); h_b[i] = static_cast<float>(rand())/static_cast<float>(RAND_MAX); }

    cl_platform_id platform_id;
    cl_device_id device_id;
    cl_int err;

    clGetPlatformIDs(1, &platform_id, NULL);
    clGetDeviceIDs(platform_id, CL_DEVICE_TYPE_GPU, 1, &device_id, NULL);
    
    cl_context context = clCreateContext(NULL, 1, &device_id, NULL, NULL, &err);
    cl_command_queue queue = clCreateCommandQueueWithProperties(context, device_id, 0, &err);

    cl_program program = clCreateProgramWithSource(context, 1, &kernelSource, NULL, &err);
    clBuildProgram(program, 1, &device_id, NULL, NULL, NULL);
    cl_kernel kernel = clCreateKernel(program, "matrix_multiply_kernel", &err);

    cl_mem d_a = clCreateBuffer(context, CL_MEM_READ_ONLY | CL_MEM_COPY_HOST_PTR, sizeof(float) * N * N, h_a.data(), &err);
    cl_mem d_b = clCreateBuffer(context, CL_MEM_READ_ONLY | CL_MEM_COPY_HOST_PTR, sizeof(float) * N * N, h_b.data(), &err);
    cl_mem d_c = clCreateBuffer(context, CL_MEM_WRITE_ONLY, sizeof(float) * N * N, NULL, &err);

    clSetKernelArg(kernel, 0, sizeof(cl_mem), &d_a);
    clSetKernelArg(kernel, 1, sizeof(cl_mem), &d_b);
    clSetKernelArg(kernel, 2, sizeof(cl_mem), &d_c);
    clSetKernelArg(kernel, 3, sizeof(int), &N);

    size_t global_item_size[2] = {(size_t)N, (size_t)N};
    
    auto start = std::chrono::high_resolution_clock::now();
    clEnqueueNDRangeKernel(queue, kernel, 2, NULL, global_item_size, NULL, 0, NULL, NULL);
    clFinish(queue);
    auto stop = std::chrono::high_resolution_clock::now();

    auto duration = std::chrono::duration_cast<std::chrono::milliseconds>(stop - start);
    std::cout << "iGPU Execution Time: " << duration.count() << " ms" << std::endl;
    
    clReleaseMemObject(d_a);
    clReleaseMemObject(d_b);
    clReleaseMemObject(d_c);
    clReleaseProgram(program);
    clReleaseKernel(kernel);
    clReleaseCommandQueue(queue);
    clReleaseContext(context);

    return 0;
}
\end{lstlisting}

\end{document}